\documentstyle[11pt,newpasp,twoside,epsf]{article}

\begin{document}
\def\Vd{V_{\rm disk}}
\def\taud{\tau_{\rm disk}}
\def\kms{{\rm km\,s^{-1}}}

\title{The Evolution of Galaxy Clustering in Hierarchical Models}
\author{Shaun Cole, Andrew Benson, Carlton Baugh, Cedric Lacey and 
Carlos Frenk.}
\affil{Department of Physics, University of Durham, South Road, Durham,
DH1 3LE U.K.}

\begin{abstract}
The main ingredients of recent semi-analytic models of galaxy
formation are summarised. We present predictions for the galaxy
clustering properties of a well specified $\Lambda$CDM model whose
parameters are constrained by observed local galaxy
properties. We present preliminary predictions for evolution of
clustering that can be probed with deep pencil beam surveys.
\vskip -1.0 truecm
\end{abstract}


\section{Introduction}

Observations now probe the properties of galaxy populations over
a large fraction of the age of the universe
(e.g. Steidel et al. 1996; Ellis et al. 1996; Lilly et al. 1996;
Adelberger et al. 1998).  Furthermore, we can look forward to a much
more detailed study of the high redshift universe with the many
instruments soon to be commissioned on the growing generation of new
8-m class telescopes. Over the lookback time  probed by these observations
the conventional cold dark matter dominated models of structure
formation predict very strong evolution of the distribution of dark matter. 
It is therefore the job of 
theorists to progress to the stage where these observations can be 
interpreted and further properties predicted within the framework 
of the hierarchical evolution of dark matter halos.

The main approaches that have been taken towards this goal can be
divided into two classes.  The first, direct simulation, involves
solving explicitly the gravitational and hydrodynamical equations in
the expanding universe using numerical N-body techniques (e.g. Katz et
al. 1992; Evrard et al. 1994; Navarro \& Steinmetz 1997; Frenk et
al. 1999; Pearce et al. 1999).  The second approach, now commonly
known as ``semi-analytic modelling of galaxy formation'', calculates
the evolution of the baryonic component using simple analytic models,
and uses a Monte-Carlo technique to generate merger trees that
describe the hierarchical growth of dark matter halos.

The two modelling techniques have complementary strengths. The major
advantage of direct simulations is that the dynamics of the cooling
gas are calculated in full generality, without the need for
simplifying assumptions. The main disadvantage is that even with the
best codes and fastest computers available, the attainable
resolution is still some orders of magnitude below that required to
fully resolve the formation and internal structure of individual galaxies in
cosmological volumes. In addition, a phenomenological model, similar
to that employed in semi-analytic models, is required to include star
formation and feedback processes.  Semi-analytic models do not suffer 
from such resolution limitations. Their
major disadvantage is the need for simplifying assumptions in the
calculation of gas properties, such as spherical symmetry which is
imposed to estimating the cooling rate of halo gas.  An important advantage of
semi-analytic models is their flexibility, which allows the effects of
varying assumptions or parameter choices to be readily investigated
and makes it possible to calculate a wide range of observable galaxy
properties, such as luminosities, sizes, mass-to-light
ratios, bulge-to-disk ratios, circular velocities, etc.

 Semi-analytic models of galaxy formation based on Monte-Carlo methods
for generating halo merger trees were pioneered by two groups, one now
based in Munich (e.g. Kauffmann, White \& Guiderdoni 1993; Kauffmann
\& Charlot 1994; Kauffmann 1995a,b; Diaferio et al. 1999) and the
other in Durham (e.g. Cole et al. 1994; Baugh et al. 1998; 
Governato et al. 1998; Benson et al. 1999a,b; Cole et al. 1999). 
There is now a
third, well established independent group (Somerville \& Primack 1999;
Somerville \& Kolatt 1999; Somerville et al. 1999) and the field
continues to grow with interesting variants being developed, for
example, by Roukema et al. (1997), Avila-Reese \& Firmani (1998), Wu,
Fabian \& Nulsen (1998) and van Kampen, Jimenez \& Peacock (1999).
The numerous contributions of these groups to this meeting are an
indication of the versatility and usefulness 
of this approach to modelling galaxy formation.

There are typically a large number of differences between the detailed
assumptions made in any two of the above models. Many of these relate
to the prescription for generating the halo merger trees. These
differences typically have little effect on model predictions. Also
one can expect this aspect of the various approaches to converge,
because in each case the models are attempting to emulate the
evolution of dark matter halos seen in high resolution N-body
simulations.  The most important differences relate
to assumptions regarding star formation, e.g. the importance of merger
induced bursts and the manner in which stellar feedback operates.  In
this respect all the models are, inevitably, oversimplified and
probably the best way forward is to confront the models continually 
with ever more detailed and accurate data.  The great value of the
semi-analytic approach is its ability to address a wide range of
observational data from galaxy luminosity functions, colour and
metallicity distributions to clustering statistics within a single
coherent model.  We have found this to be a particular strength of the
models as it often allows robust predictions to be made despite the
intrinsic uncertainty in the physical processes that are being
modelled.

In the remainder of this article we briefly describe 
the latest Durham  model and use it to illustrate the main processes 
that are incorporated in semi-analytic models of galaxy formation. 
We then present a comprehensive set of results for the 
galaxy clustering properties predicted by this model for a $\Lambda$CDM
($\Omega_0=0.3$, $\Lambda_0=0.7$) cosmology, after the
model is constrained to
reproduce the the bright end of the observed galaxy luminosity function.

\section{The Model}

A full description of the current Durham  semi-analytic 
galaxy formation model, complete with an exploration of how the
predictions depend on parameter variations and how they compare to 
observational data, can be found in Cole et al. (1999). Here
we simply describe the main features of the model.

\subsection{Merger Trees}

We use a simple new Monte-Carlo algorithm to generate merger trees
that describes the formation paths of randomly selected dark matter
halos. Our algorithm is based directly on the analytic expression for
halo merger rates derived by Lacey \& Cole (1993).  The algorithm
enables the merger process to be followed with high time resolution,
as timesteps are not imposed on the tree but rather are controlled
directly by the frequency of mergers.  Also, there is no quantization
of the masses of the halos.


\subsection{Halo Structure and Gas Cooling}

We assume that the dark matter in virialized halos is well described
by the NFW density profile (Navarro, Frenk \& White 1997).  We further
assume that any diffuse gas present during a halo merger is 
shock-heated to the virial temperature of the halo. The density profile we
adopt for the hot gas is less centrally concentrated than that of the
dark matter and is chosen to be in agreement with the results of high
resolution simulations of non-radiative gas (e.g. Frenk et al. 1999).  We
estimate the fraction of gas that can cool in a halo by computing the
radius at which the radiative cooling time of the gas equals the age
of the halo. The gas that cools is assumed to conserve angular
momentum and settle into a rotationally supported disk. Thus, the
initial angular momentum of the halo, which we assign using the well
characterised distribution of spin parameters found for halos in
N-body simulations, determines the size of the resulting galaxy
disk. In computing the size of the disk we also take account of the
contraction of the inner part of the halo caused by the gravity of the disk.

\subsection{Star Formation and Feedback}

The processes of star formation and stellar feedback are the most uncertain to
model. We adopt a flexible approach in which the star formation rate in the
disk of cold gas is given by
$\dot M_{\star} = M_{\rm cold}/ \tau_\star$, 
with the timescale $\tau_\star$ parameterized as
\begin{equation}
\tau_\star = \epsilon_\star^{-1} \,\taud \, (\Vd/200 \, \kms)^{\alpha_\star}. 
\end{equation}
We also adopt a feedback model in which for every solar mass of stars formed,
\begin{equation}
	\beta = (\Vd/V_{\rm hot})^{-\alpha_{\rm hot}} 
\end{equation}
solar masses of gas are assumed to be reheated and ejected from the disk as a
result of energy input from young stars and supernovae.  In these
formulae, $\taud$ and $\Vd$ are the dynamical time and circular
velocity of the disk; $\epsilon_\star$, $\alpha_\star$, 
$\alpha_{\rm hot}$ and $V_{\rm hot}$ are the model parameters.

\subsection{Galaxy Mergers}

Mergers between galaxies can occur, subsequent to the merger of their
dark matter halos, if dynamical friction causes the orbits of the
galaxies to decay.  The result of a merger depends on the mass ratio
of the merging galaxies.  If they are comparable, $M_{\rm
smaller}>f_{\rm ellip}M_{\rm larger}$, then the merger is said to be
violent and results in the formation of a spheroid. At this point any
cold gas present in the merger is assumed to undergo a burst of star
formation, with a timescale equal to the dynamical time of the forming
spheroid and with feedback estimated using equation (2), but with
the circular velocity of the spheroid replacing that of the disk.
The size of the resulting spheroid is estimated assuming energy
conservation in the merger (once dynamical friction has eroded the
orbits to the point where the galaxies interpenetrate) and the virial
theorem.  For minor mergers, $M_{\rm smaller}<f_{\rm ellip}M_{\rm
larger}$, we assume the cold gas is accreted by the disk and
the stars by the bulge of the larger galaxy.

\subsection{Stellar Population Synthesis and Dust}

To convert the calculated star formation histories of each galaxy into
observable luminosities and colours we use the stellar population
synthesis model of Bruzual \& Charlot (1993,1999) and, in addition, the
3-dimensional dust model of Ferrara et al. (1999).  For the former we
adopt the IMF of the solar neighbourhood as parameterized by Kennicutt
(1983) and for the latter we adopt their Milky Way extinction law and
assume that the dust/gas ratio in the cold gas disk scales with
metallicity.

\subsection{Galaxy Clustering}

Given a list of halo masses, at the present day or at some redshift $z$,
the above model can be used to determine the number, luminosity
and other properties of the galaxies that inhabit them. 
It is then straightforward to compute the amplitude of their
correlation function on large scales using the formalism of
Mo \& White (1996). 
Here we use a more direct approach, 
that of using the positions of halos from an N-body simulation,
as this enables the correlation function to be studied down to smaller
scales and allows other aspects of galaxy clustering to be investigated.
Details of our procedure can be found in Benson et al. (1999a).

\begin{figure}
\plotone{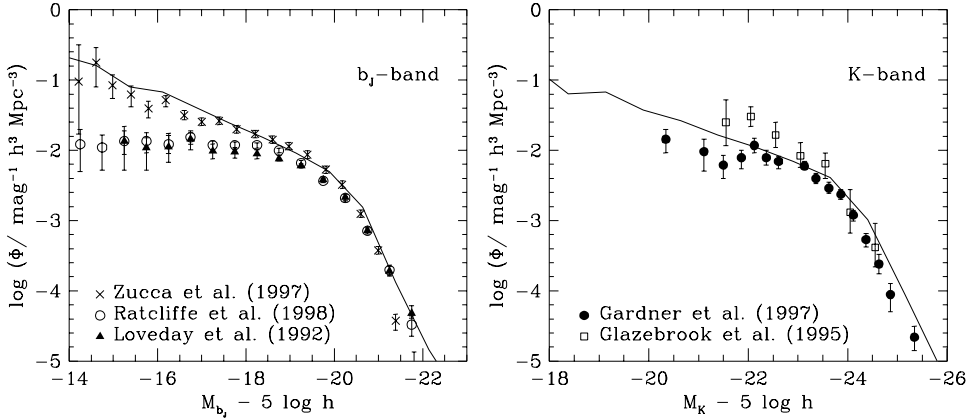}
\caption{The ${\rm b_{\rm J}}$ and K-band luminosity functions of
our $\Lambda$CDM model compared to a variety of observational
estimates (symbols).} 
\vskip -0.5 truecm
\end{figure}

\begin{figure}
\plottwo{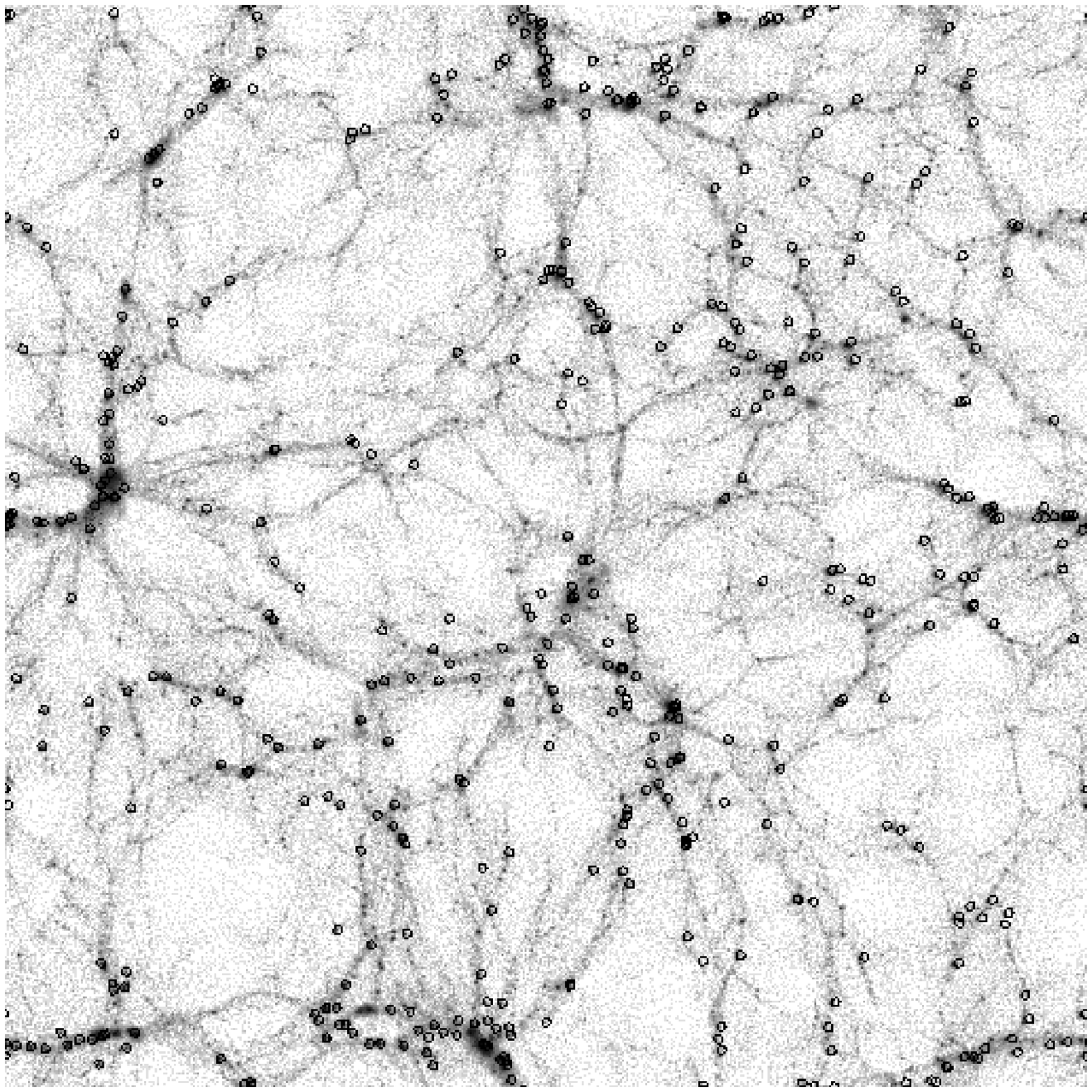}{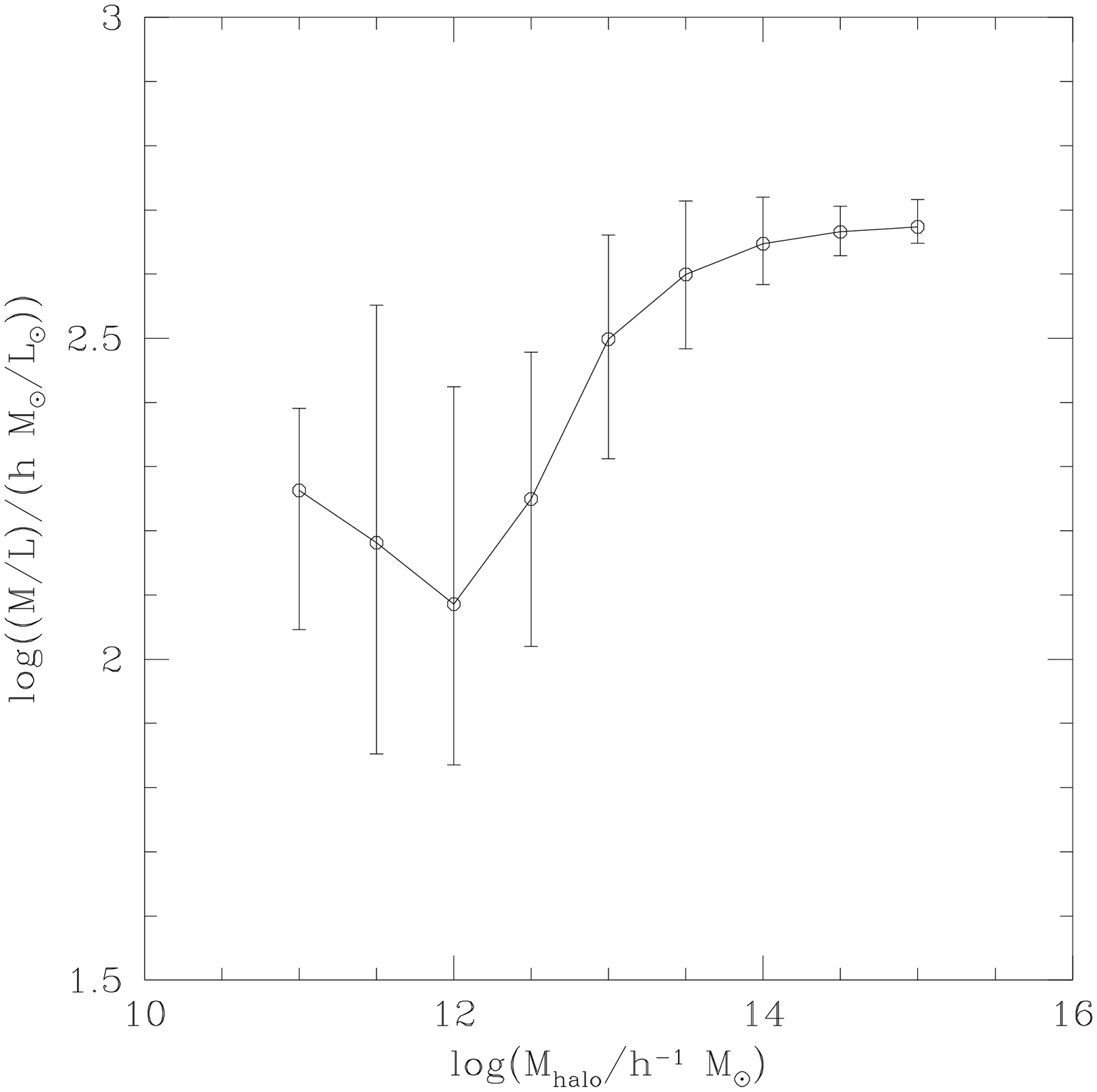}
\caption{The left hand panel shows a
 $141 \times 141 \times 8 h^{-3} {\rm Mpc}^3$
slice of the $\Lambda$CDM dark matter simulation with the positions
of galaxies brighter than ${\rm M_B}-5 \log h=-19.5$ overlaid.
The right hand panel shows the variation of the total
mass-to-light ratio of halos as a function of mass.}
\vskip -0.5 truecm
\end{figure}

\section{Observational Constraints}

The number of parameters  in our  galaxy formation model
is not small and the intrinsic range of behaviour
that the model can produce is large. This is inevitable as
galaxy formation, at the very least, involves all the processes
which are included in the model and quite possibly others as well.
Thus, progress can only be made if a set of constraints is applied 
to fix model parameters.
Our approach, set out in detail in Cole et al. (1999), is to fix
these parameters using the observed properties
of the local galaxy population, e.g. the B and K-band 
luminosity functions, the slope of the Tully-Fisher
relation and the gas fractions and metallicities of disk galaxies.
The result of applying these constraints is a well specified
model whose properties can be examined and critically compared to
other observational data such as galaxy clustering or high redshift
observations.

 Fig.~1 shows the ${\rm b_{\rm J}}$ and K-band luminosity functions of
a $\Lambda$CDM model ($\Omega_0=0.3$, $\Lambda_0=0.7$) 
constrained in this way. It turns out that the predicted low redshift
galaxy clustering is insensitive to changes in the model parameters
provided only that the model is constrained to produce a reasonable
match to the bright end of the luminosity function (Benson et
al. 1999a).

\section{Results}

   We now look at a variety of clustering properties that we predict
for the constrained $\Lambda$CDM model and compare them with available
observational data. The N-body simulation used to assign positions to
our galaxies is the $\Lambda$CDM ``GIF'' simulation carried out by the
Virgo consortium.

These same simulations have been analyzed in great detail by the
Munich group, with results presented at this meeting and in Kauffmann
et al. (1999a,b) and Diaferio et al. (1999).  Their approach is more
sophisticated than ours in that they extract the halo merger trees
directly from the N-body simulations and are able to follow individual
galaxies in the simulation from one epoch to another.  However, these
differences in approach do not significantly affect the predictions of
the clustering and kinematic properties of the galaxy populations. In
particular, with Antonaldo Diaferio, we were able to verify that if we
use our algorithm to assign galaxies positions, but start with the
Munich group's list of galaxies as a function of halo mass, we recover
very similar results to those reported in Diaferio et al. (1999).
These tests are discussed in Benson et al. (1999b), where we conclude
that in the few cases where significant differences exist between our
results and those of the Munich group, they are largely a result of
the differing constraints that have been applied to the models.

\begin{figure}
\plottwo{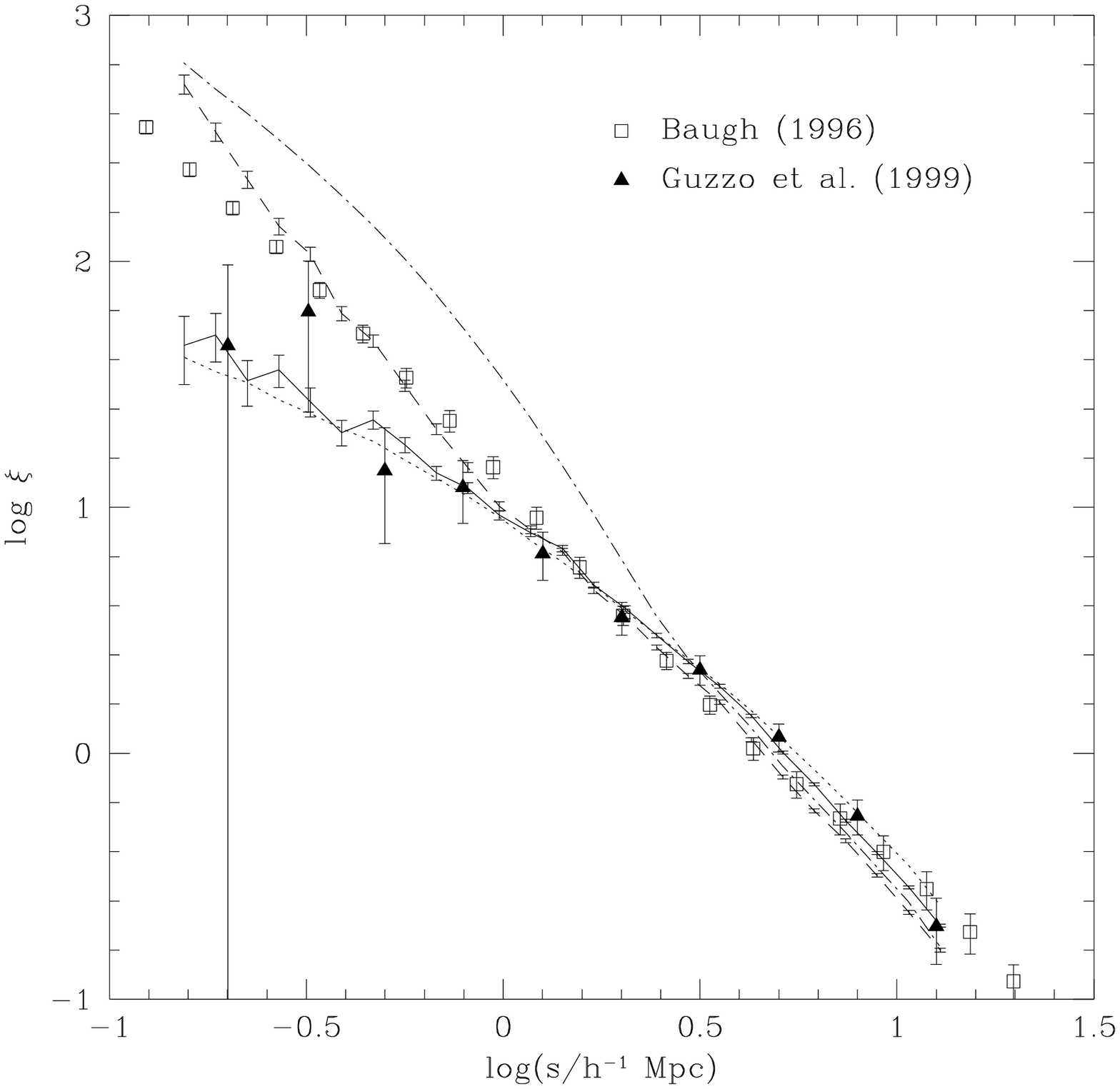}{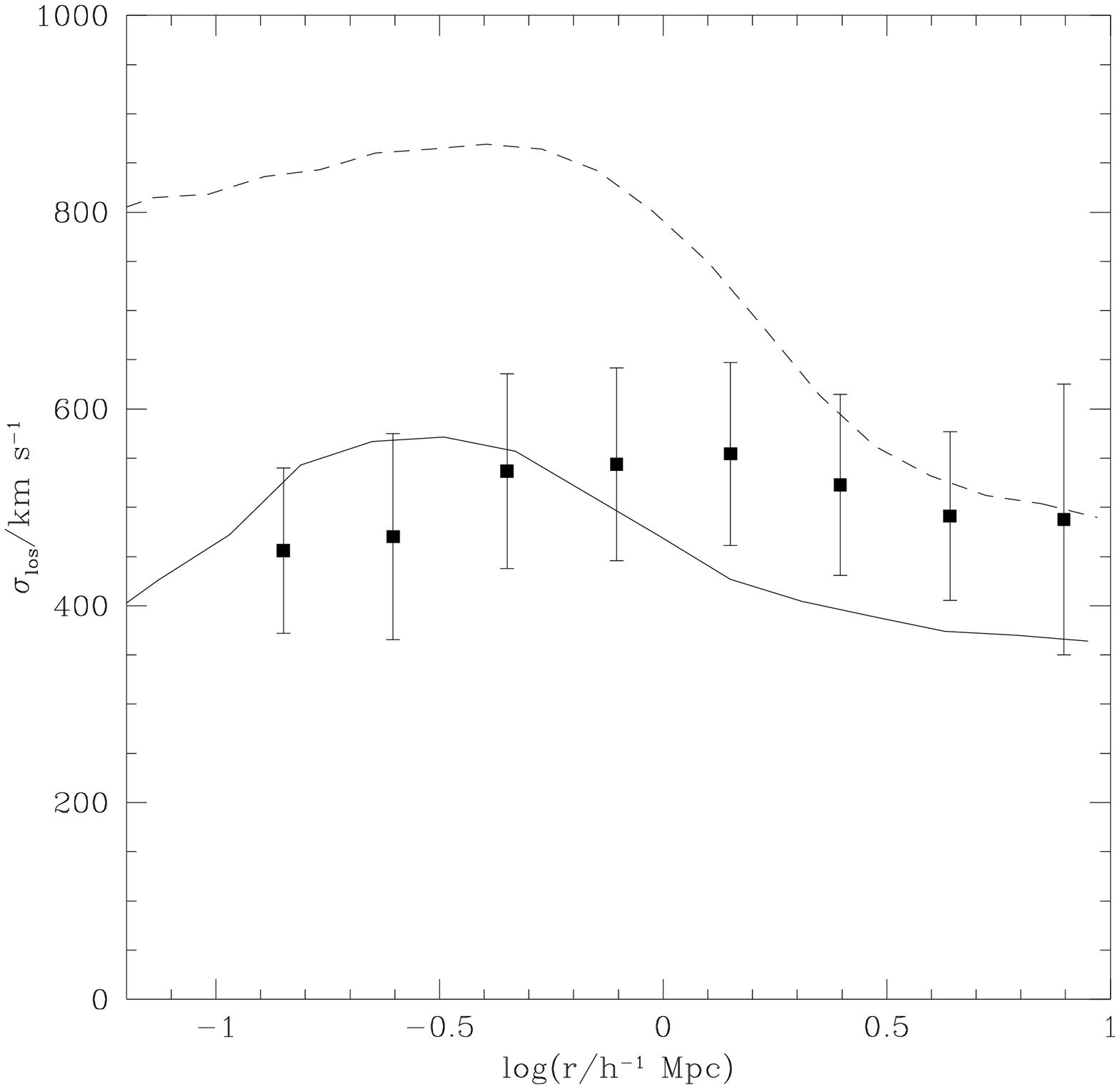}
\caption{The left hand panel shows real/redshift space 
correlations functions of the dark matter (dot-dashed/dotted) and galaxies 
(dashed/solid) and also observational estimates of the 
real and redshift space galaxy correlation function.
The right panel shows the pairwise velocity dispersion, $\sigma_{\rm los}$, 
of dark matter (broken line) and galaxies (solid line) and an 
observational estimate from the LCRS (Jing, Mo \& Borner 1998).}
\vskip -0.5 truecm
\end{figure}

\begin{figure}
\plotfiddle{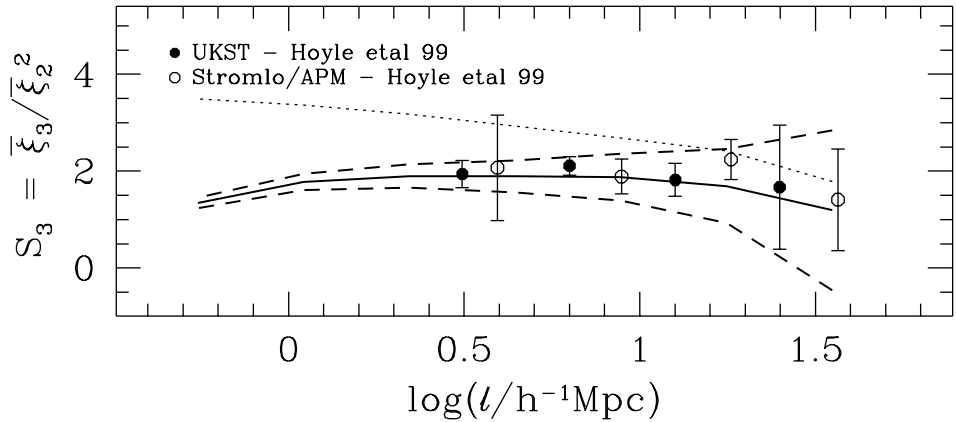}{5.0 truecm}{0}{75}{75}{-240}{-150}
\caption{The skewness, $S_3$, of the distribution of counts in cubical cells
in redshift space as a function cell size. The dotted line shows $S_3$ 
for the dark matter and the solid line $S_3$ for the galaxies. 
The dashed lines show the 1-$\sigma$ uncertainty in this estimate. 
The points with errorbars show two recent observational estimates.}
\vskip -0.5 truecm
\end{figure}

\subsection{Low Redshift Clustering}

We start, in Fig.~2a, by showing a slice through the N-body simulation
with the positions of galaxies superimposed on the dark matter
distribution.  It is worth noting that the way in which galaxies trace
the dark matter is non-trivial with galaxies avoiding the large
underdense regions and concentrating in filaments and clusters.  In
our model this distribution is entirely determined by the combination
of the dark matter distribution and the distribution of the number of
galaxies within halos as a function of halo mass. One representation
of this distribution is Fig.~2b, which shows the variation of total
mass-to-light ratio of halos as a function of halo mass, with the
errorbars indicating the 10 and 90~centiles of this distribution. This
dependence is produced naturally by the physics incorporated
into the semi-analytic model.  Galaxy formation is most efficient (M/L
lowest) in intermediate mass halos. The efficiency is reduced in low
mass halos due to feedback and in the most massive halos due to long
cooling times.  The low efficiency in low mass halos leads to the
production of large voids in the galaxy distribution, while the
inefficiency at high masses leads to an anti-bias in small scale clustering,
as shown by the real-space correlation functions of Fig.~3a.
In fact the real-space galaxy correlation function has a nearly power
law form in quite good agreement with that observed.  
The underrepresentation of galaxies in clusters also leads to a  
reduced pairwise velocity dispersion (Fig.~3b) which is clsoe to
the observed value.
Strangely, the different dark matter and galaxy
peculiar velocities act to produce very similar redshift space correlation
functions for both components and these match well the
observed redshift-space galaxy correlation function.

Fig.~4 compares the skewness, $S_3$, in redshift space as a 
function of cell size with two recent observational estimates.
The skewness of the model galaxy distribution is substantially less
than that of the dark matter and in remarkably good agreement with the
observed values.

\begin{figure}
\plotfiddle{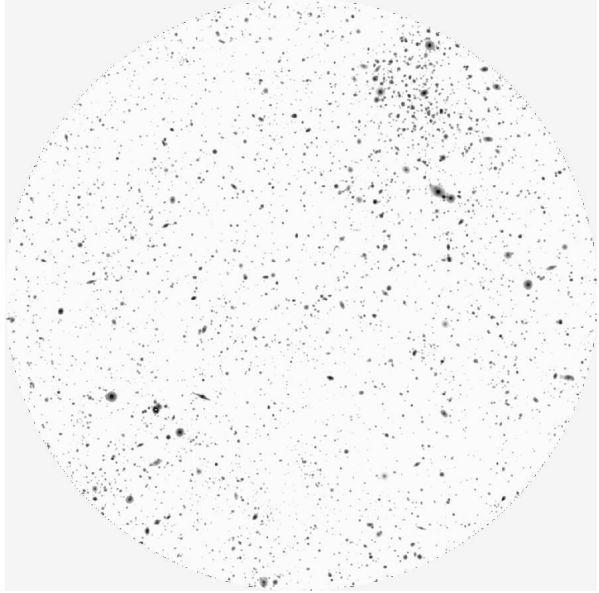}{7 truecm}{0}{25}{25}{-80}{10}
\caption{A mock image of a 18 arc minute diameter field to a limiting
magnitude of R$=24$ constructed from a simulated light-cone.}
\vskip -0.5 truecm
\end{figure}

\subsection{The Evolution of galaxy clustering}

A strong test of the models will come from comparing their
predictions against high redshift observations. One highly successful
comparison has already been made. In Baugh et al. (1998) and
Governato et al. (1998) firm predictions for the Lyman-break galaxy
correlation function were made. These later proved to agree remarkably
well with observations (Adelberger et al. 1998). 
Unfortunately the predictions for
both low and high $\Omega_0$ models are very similar and so the
observations do not discriminate between cosmological models.

In order to make detailed comparisons of the galaxy formation models with
the ever increasing quantity of high redshift data we are
developing techniques to simulate deep pencil beam surveys
and accurately match observational selection criteria.
Fig.~5 shows a simulated deep R-band image constructed using our technique
of outputting a light-cone from an evolving N-body simulation and then
using the semi-analytic galaxy formation model to populate its halos
with galaxies. Given the limited resolution and volume of the simulation 
used to construct these prototypes one should be somewhat cautious to avoid
over-interpreting their predictions. Nevertheless it is interesting
to make a preliminary comparison with some recent observations.

Fig.~6 compares predictions for the clustering amplitude at one degree
as a function of limiting K-magnitude and for redshift slices. The
model predictions are based on an ensemble of simulated light cones with
the errorbars indicating the rms scatter. The model agrees well with
the K-band observations of McCracken (1999). The model predicts very little
variation of clustering amplitude with redshift, whereas Magliocchetti
\& Maddox (1999) find quite a strong trend. However, the errorbars are large
and larger area surveys will be required to critically test the models.

\begin{figure}
\plottwo{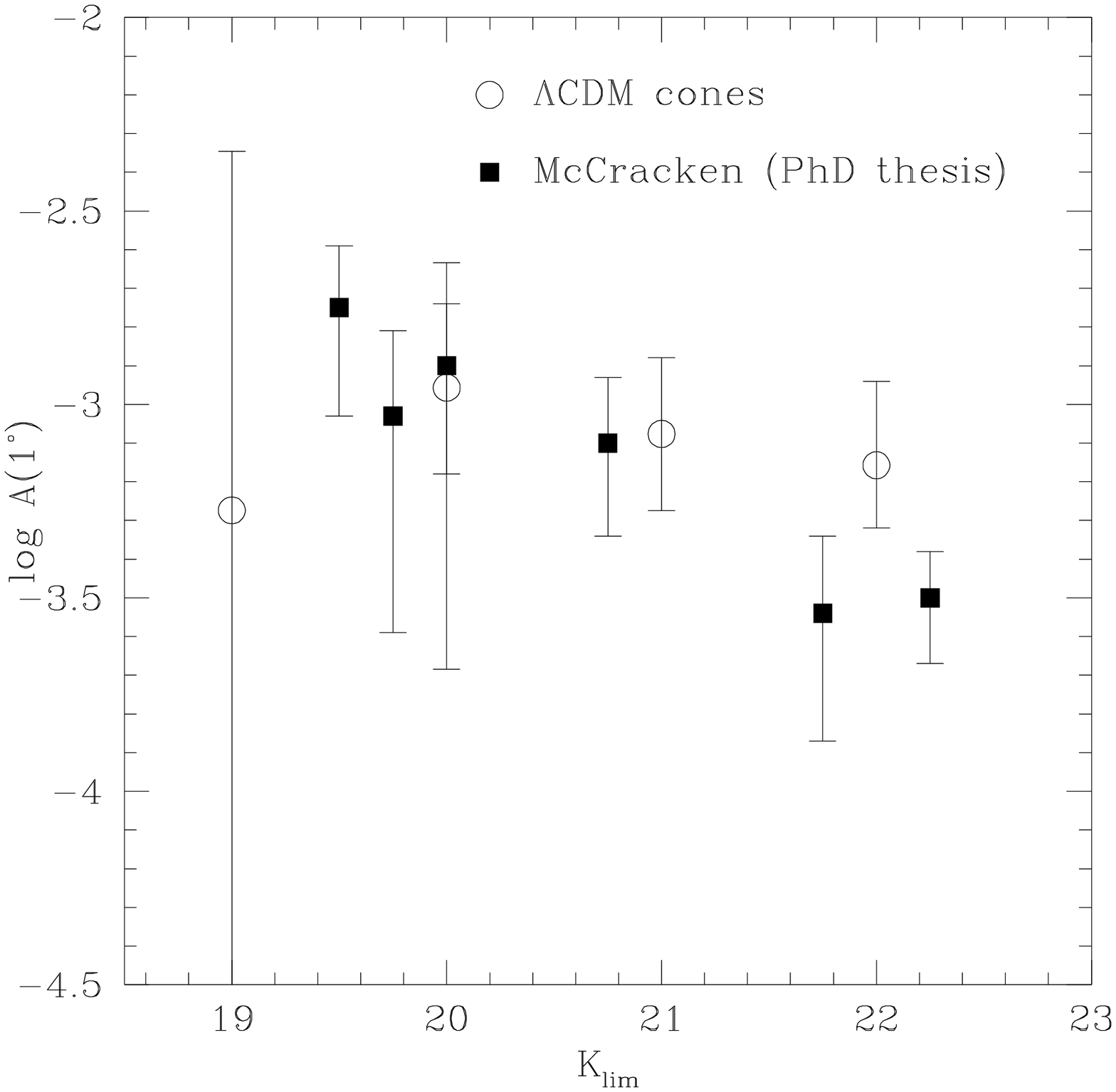}{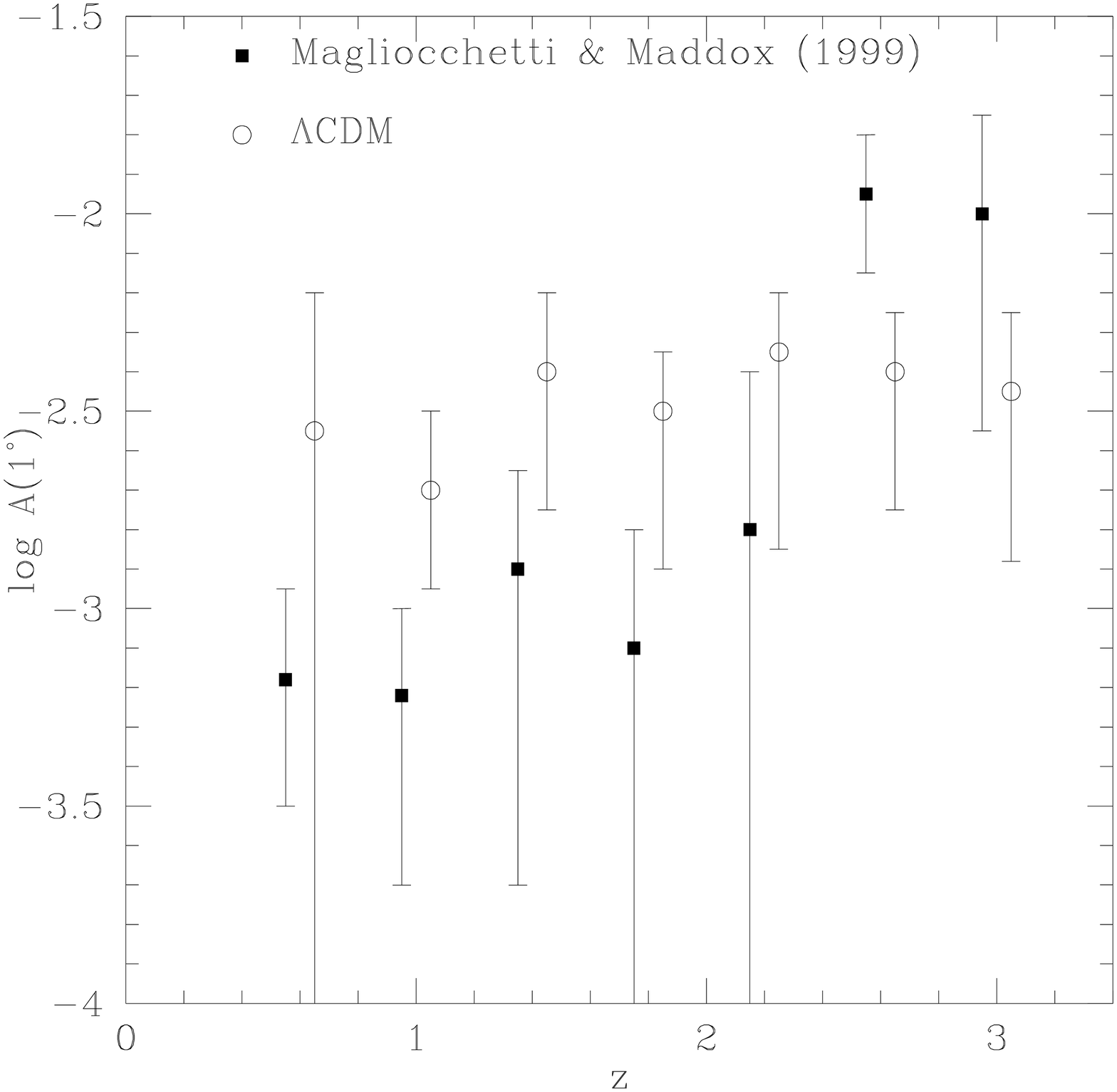}
\caption{The model prediction for the dependence of the
amplitude at one degree of the angular correlation function $w(\theta)$
versus a) limiting K-band magnitude and b) redshift.
In each case the selection criteria were chosen to mimic those
of the observational dataset with which they are compared.  }
\vskip -0.5 truecm
\end{figure}

\section{Conclusions}

Physically motivated semi-analytic models of galaxy formation
which, importantly, include the evolution of structure,
provide a framework in which very diverse properties
of galaxies can be modelled and understood. Using a subset
of the locally observed galaxy properties  these
models can be constrained and then employed to
make useful predictions. Such predictions include
all aspects of galaxy evolution and also galaxy clustering.

We have only examined the predicted clustering properties
for a couple of cosmological models (Benson et al. 1999a). However,
it is intriguing that the results of the $\Lambda$CDM model presented 
here appear to match galaxy clustering data remarkably well
and significantly better than a $\tau$CDM model.

We have invested in the technology necessary to 
extend the model predictions to high redshift by simulating
deep pencil beams. Soon, such data will provide very interesting 
tests of galaxy formation models.

\end{document}